# Optical wireless link between a nanoscale antenna and a transducing rectenna

Arindam Dasgupta, Marie-Maxime Mennemanteuil, Mickaël Buret, Nicolas Cazier, Gérard Colas-des-Francs and Alexandre Bouhelier[*]

*Laboratoire Interdisciplinaire Carnot de Bourgogne, CNRS UMR 6303, Université de Bourgogne Franche-Comté, 9 Avenue A. Savary, 21000 Dijon, France*

[*]e-mail : alexandre.bouhelier@u-bourgogne.fr

**Initiated as a cable-replacement solution, short-range wireless power transfer has rapidly become ubiquitous in the development of modern high-data throughput networking in centimeter to meter accessibility range[1]. Wireless technology is now penetrating a higher level of system integration for chip-to-chip and on-chip radiofrequency interconnects[2]. However, standard CMOS integrated millimeter-wave antennas have typical size commensurable with the operating wavelength, and are thus an unrealistic solution for downsizing transmitters and receivers to the micrometer and nanometer scale. In this letter, we demonstrate a light-in and electrical-signal-out, on-chip wireless near infrared link between a 200 nm optical antenna and a sub-nanometer rectifying antenna converting the transmitted optical energy into direct current (d.c.). The co-integration of subwavelength optical functional devices with an electronic transduction offers a disruptive solution to interface photons and electrons at the nanoscale for on-chip wireless optical interconnects.**

Conventional radiowave and microwave antennas operate a bilateral energy conversion between electrical signals and electromagnetic radiations. Integration of these antennas to modern consumer electronics is thus widely adopted for long-range and short-range data transfer. Vivid examples are communication between mobile devices and remote biosensors for healthcare providers[3,4]. Similar sub-wavelength interfacing of optics and electronics is envisioned as an archetype to reduce the size of communication devices while maintaining their necessary speed and bandwidth[5,6]. This is a daunting task as state-of-the-art electronics is unable to respond to the fast alternating fields associated to optical frequencies and the size of photonic components remains orders of magnitude larger than their electronic counterparts. Until now, optical antennas have been mainly limited to interfacing near-field and far-field radiation by tailoring the momentum of light[7].

In 2010, A. Alù and N. Engheta theoretically proposed an optical wireless channel between two nano-scale antennas[8]. The idea was partially demonstrated by beaming either an excitation signal towards a distant luminescent receiver[9] or by the mediation of surface plasmons[10]. Despite an optimization with highly directive antennas[11,12], the link operates on the basis of a light-in and light-out configuration without transduction of the transferred optical energy to an electronic signal.

Recent progresses show that tunneling nonlinearity in the conduction of an atomic scale tunnel junction can rectify the plasmonic response at optical frequencies to d.c. current[13-17]. Immediately, these rectifying antennas or rectennas, appear as essential ultrafast devices for merging optics and electronics at the nanoscale. Here, we exploit these functionalities on a single platform to realize an optical wireless power transmission between an illuminated nanoscale dipolar antenna and a rectenna.

Figure 1a illustrates the line-of-sight optical channel presented here. The transmitter is a laser-illuminated gold nanodisk acting as an optical dipole antenna. The radiation broadcasted by the antenna is detected and converted to a d.c. current by an electrically biased rectenna. An optical image of the units is shown in Fig. 1b. The Au electrodes powering the rectenna feed-gap and a series of optical antennas (black dashed box) are readily seen with a dark contrast. The tunneling feedgap of the rectenna is formed by electromigration[18]. Figure 1c is a scanning electron microscopy (SEM) image of the white dotted box in Fig. 1b showing the in-plane tunneling junction together with an adjacent optical antenna. The electrical characterization confirms the presence of a tunneling transport (see methods section). The optical antennas have a fixed diameter of 230±10 nm and are resonant with the excitation wavelength[19]. The devices are immersed in a refractive index matching oil to operate the link in a homogenous environment.

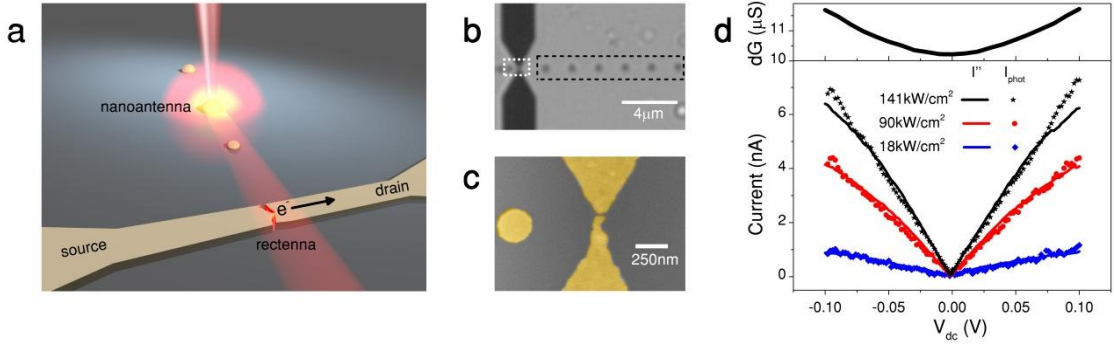

**Figure 1| Concept and the device characterization. a**, Schematic illustration of the optical wireless transducing link: an optical antenna is excited by a focused laser beam. The energy radiated by the transmitter antenna is detected and converted to an electrical signal by a distant electrically biased receiving rectenna. **b**, Optical transmission image of the functional units. The white dotted box highlights the feed-gap of the rectenna and the black dotted frame contains a series of transmitter optical antennas placed at different distances to the rectifying gap. **c**, SEM image of the region highlighted by the white dotted rectangle in b featuring a rectifying nanoscale feed-gap between two gold electrodes. **d**, Plot of the variation of the amplitude of the rectified photocurrent $I_{phot}$ (data points) and of the current proportional to the nonlinearity of junction's conductance $I''=1/4(V_{ac}^2 \partial^2 I / \partial V^2)$ (line plots) as a function of applied bias $V_{dc}$ and for three direct excitation intensities of the rectenna illuminated by a focused 785 nm laser. The shared trends between $I_{photo}$ and $I''$ confirm an optical rectification mechanism. The upper graph shows the evolution of the differential conductance dG of the tunnel feed-gap with $V_{dc}$.

Let us assume a tunnel junction where the conduction mechanism remains the same over a pulsation range $2\omega_{ac}$. The junction produces a d.c. current $I(V_{dc})$ when a bias of $V=V_{dc}$ is applied between the two metal leads. If a small a.c. bias $V_{ac}$ of frequency $\omega_{ac}$ is superposed to $V_{dc}$, the total current through the junction can be expressed by a Taylor's expansion[20]

$$I = I(V_{dc}) + \left.\frac{\partial I}{\partial V}\right|_{V_{dc}} V_{qc} \cos \omega_{ac} t + 1/2 \left.\frac{\partial^2 I}{\partial V^2}\right|_{V_{dc}} V_{ac}^2 \cos^2 \omega_{ac} t + \ldots$$

$$\approx \left[ I(V_{dc}) + \frac{1}{4}\left.\frac{\partial^2 I}{\partial V^2}\right|_{V_{dc}} V_{ac}^2 \right] + \left.\frac{\partial I}{\partial V}\right|_{V_{dc}} V_{ac} \cos \omega_{ac} t + \frac{1}{4}\left.\frac{\partial^2 I}{\partial V^2}\right|_{V_{dc}} V_{ac}^2 \cos 2\omega_{ac} t + \ldots \quad (1)$$

The time-independent term in Eq. 1 indicates the presence of an additional rectified current $I''$, which is proportional to the nonlinearity of the conductance $\partial^2 I / \partial V^2$ and $V_{ac}^2$, $I''=1/4 V_{ac}^2 (\partial^2 I / \partial V^2)$.

To describe optical rectification, a quantum mechanical treatment is generally used[21]. When the rectenna is illuminated with light of energy $\hbar\omega$, the response builds up an a.c. voltage $V_{opt}$ of pulsation $\omega$ across the junction. This optical potential triggers a photon-assisted tunneling of electrons to produce a d.c. photocurrent in addition to the $I(V_{dc})$. If $eV_{opt} << \hbar\omega$, the rectified d.c. photocurrent is given by[22,23]

$$I_{phot} = I(V_{dc}, V_{opt}, \omega) - I(V_{dc}) = \frac{1}{4} V_{opt}^2 \left[ \frac{I(V_{dc} + \hbar\omega/e) - 2I(V_{dc}) + I(V_{dc} - \hbar\omega/e)}{(\hbar\omega/e)^2} \right] \quad (2)$$

For gold and for an excitation energy $\hbar\omega < 2$ eV, the tunneling transmission remains smooth within the range $E_F \pm \hbar\omega$, where $E_F$ is the Fermi energy[13]. Therefore, Eq. 2 reduces to its classical form $I_{phot} = 1/4 V_{opt}^2 (\partial^2 I/\partial V^2)$ with $V_{opt} = V_{ac}$ and $I_{phot} = I''$.[13,24]

We use a low frequency a.c. voltage and lock-in detection to record the differential conductance $dG = \partial I/\partial V$, the current proportional to the nonlinearity $\partial^2 I/\partial V^2$ and the laser-induced current $I_{phot}$[13,14,20]. The description of the complete measurement system is included in the methods section. The plot of differential conductance of the junction $dG(V_{dc})$ in the inset of Fig.1d features a zero bias conductance of about 10 μS (≈0.13 $G_0$) where $G_0$ is the quantum of conductance (77.5 μS). Using Simmons' model[25], we qualitatively estimate the gap width to be < 0.5 nm (see supplementary info). In Fig. 1d, we plot the bias dependence of $I''$ (line) and $I_{phot}$ (points) for three laser intensities directly focused on the rectenna feed-gap. For each excitation intensity, $V_{ac}$ is adjusted to obtain $I_{phot} = I''$, indicating that the optical voltage generated across the feed-gap equals the low frequency voltage applied between the electrodes[13]. From the graph it is evident that $I_{phot}$ follows $I''$ suggesting the photocurrent generated at the rectenna results from optical rectification. Additional experiments to rule out any thermal contributions are presented in supplementary information.

In the following section, we assess the operation of a wireless link when a remote optical antenna is broadcasting a signal towards this transducing rectenna. Figure 2a shows a pixel-by-pixel reconstructed photocurrent map generated by the rectenna. The laser excitation is polarized along Y axis (0°) and the sample is scanned through the focal area (step size 70 nm). $V_{dc}$ is fixed at 50 mV and the laser intensity at 353 kWcm$^{-2}$. The important conclusion drawn from the current map is the presence of a rectenna response whenever the laser excites a remote optical antenna with a range exceeding several micrometers. When the polarization is turned by 90° (X-axis), the photocurrent generated at the rectenna vanishes nearly completely (Fig. 2b). To confirm the transduction of the signal radiated by the optical antennas, we station the laser on an optical antenna 4 μm away from the rectenna (2$^{nd}$ antenna from the right in Fig. 1b) and simultaneously monitor $I_{phot}$ and $I''$ as a function of $V_{dc}$ while varying the polarization. The intensity of the laser is here 540 kWcm$^{-2}$. For all the incident polarizations, $I_{phot}$ follows $I''$ proving thus the $I_{phot}$ signal is optically rectified (see supplementary info.). The value of $V_{ac}$ for conditioning $I_{phot}=I''$ is recorded as a measure of the optically induced a.c. voltage $V_{opt}$ at the feed-gap. The evolution of $I_{phot}$ and $V_{opt}$ as a function of the incident laser polarization is plotted in figure 2c for a bias $V_{dc}$=50 mV. It is clear that rectification process is suppressed as the polarization is turned by 90°. To verify that $I_{phot}$ does not originate through any optically-induced changes of the conductance[19], we map $dI/dV$ at the frequency of the optical chopper (see supporting info). No measurable contrast can be related to the photocurrent maps presented in Fig 2a.

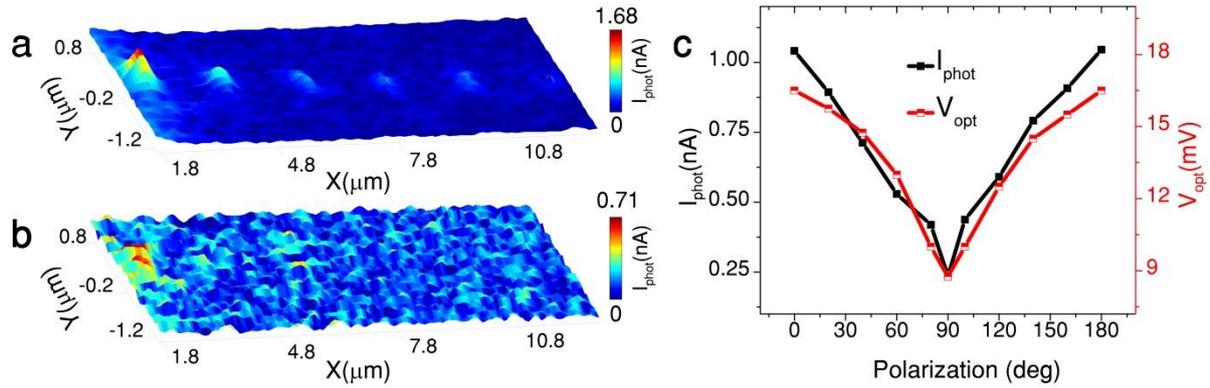

**Figure 2| Polarization response of the optical rectenna receiving an electromagnetic signal emitted from individually excited optical antennas. a,** Photocurrent map generated by the rectenna reconstructed pixel-by-pixel by scanning the optical antennas through the laser focus. The scanned region is indicated by the black dotted rectangle in Fig. 1b. The incident polarization is along the vertical direction (0°) for excitation intensity of 353 kWcm$^{-2}$. **b,** The same region for an incident polarization along the horizontal direction (90°). $V_{dc}$ is constant at 50 mV. **c,** Plot of $I_{phot}$ and $V_{opt}$ as a function of incident polarization for individual excitation of a nanodisk which is 4 μm away for an excitation laser intensity 540 kWcm$^{-2}$.

Numerical simulations based on a three-dimensional finite element method (3D-FEM) bring an understanding of the polarization dependence of the rectified signal. The rectenna is modeled by two Au triangles separated by a distance of 10 nm except at the middle where a small protrusion on the bottom electrode reduces the gap size to 0.5 nm. A 220 nm diameter Au nanodisk is placed 4 μm away from the feed-gap. The entire geometry is placed inside a homogeneous medium of refractive index 1.52. When excited by a linear polarization at 785 nm, the disk behaves as a dipolar resonant antenna radiating its characteristic two-lobe pattern perpendicularly to the incident electric field. Figures 3a and d are schematic representations picturing the dipolar radiation for two orthogonal in-plane polarizations. Figure 3b is the calculated electric field distribution plotted in logarithmic scale when the antenna is illuminated (the white circle indicates the excitation area) with a polarization along the y-axis. Clearly, the optical antenna redirects the far-field radiation towards the rectenna feed-gap. The interaction of this radiation with the tunneling gap results in a very high electric field at the junction, enhanced by a factor of approximately 18 compared to the excitation field, which is illustrated in the zoomed-in electric field map in Fig. 3c. When the dipolar radiation is broadcasted perpendicularly to the receiver (Fig. 3d), the electric field at the junction is minimal (Fig. 3e and 3f), explaining the vanishing photo-response of Fig. 2d.

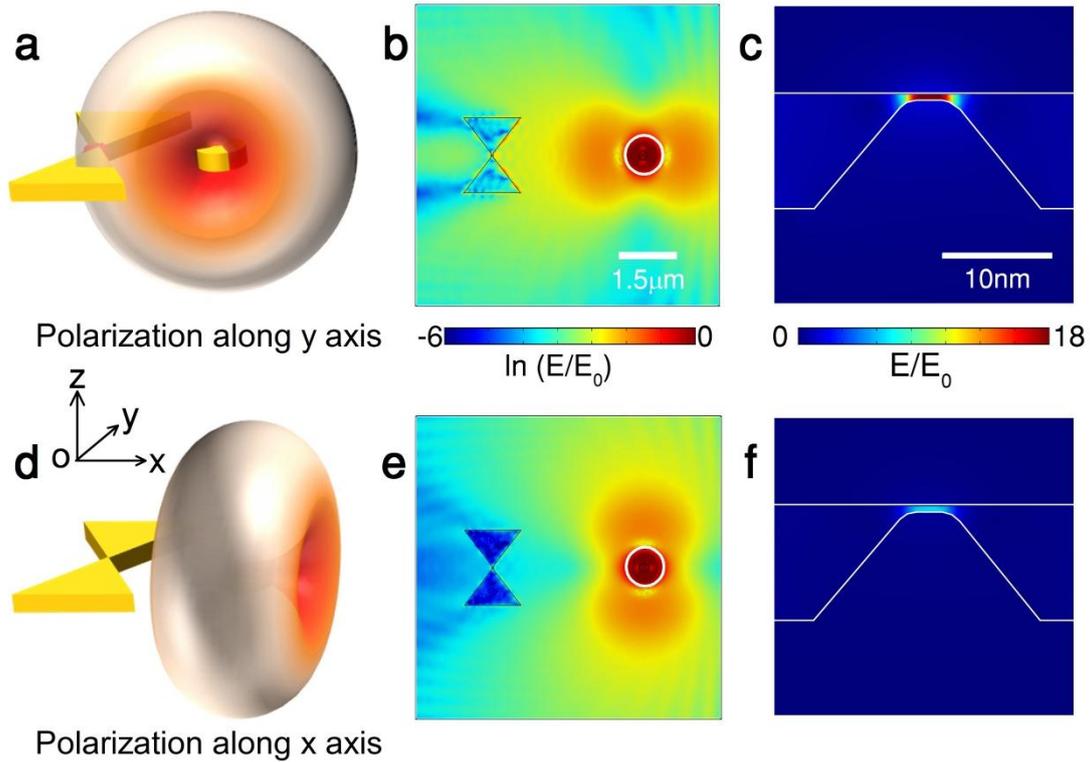

**Figure 3| Numerical simulations of the polarization dependence of the wireless power transfer. a,** Schematic of the operation principle of the wireless link when a transmitting antenna is illuminated with a laser polarized along y-axis. For this incident polarization, the dipolar radiation from the nanodisk is directed towards the receiving antenna. **b,** Calculated distribution of the electric field, plotted in log scale, for an incident polarization along y-direction. The white circle indicates the size of the excitation spot, here considered as 1 m in diameter. **c,** Zoomed image of the electric field distribution present at the feed-gap of the rectenna. **d,** Same configuration with an incident polarization along x-axis. The disk radiates in the orthogonal direction and hence the rectenna receives minimum amount of the transmitted energy. **e,** and **f,** are the electric field maps for an incident polarization along the x-axis. For clarity, the electric field in f is multiplied by a factor 5. The optical electric field created at the gap decreases drastically when the signal emitted by the antenna is not directed toward the rectifying feed-gap.

When a transmitter and a receiver constituting a wireless link are separated by a distance $d$, the received power is given by Friis equation[26],

$$P_r = \left[ \eta_r \eta_t D_r D_t \left(1-|\Gamma_r|^2\right)\left(1-|\Gamma_t|^2\right) |a_r^* a_t|^2 \frac{\lambda_{ex}}{4\pi d^2} \right] P_{fed} \quad (3)$$

$P_{fed}$ is the power fed to the transmitter, $\lambda_{ex}$ is the operational wavelength, $\eta_r$ and $\eta_t$ are the radiation efficiencies, $D_r$ and $D_t$ are the directivities, $\Gamma_r$ and $\Gamma_t$ are the reflection coefficients, $a_r$ and $a_t$ represent the polarizabilities of the receiving and the transmitting antennas, respectively. We first analyze the distance dependence of the optical wireless link. The evolution of the amplitude of the rectified current is plotted in Fig. 4a as a function of the distance $d$ separating the antenna to the receiving rectenna. $I_{phot}$ (data points) is fitted with a generic power law function $ad^b+c$ where $a$ represents the coupling strength between the two units and $c$ is the dark photocurrent of the device (≈0.32nA). The best fit gives an exponent $b=-1.8$, which is close to the expected inverse square law dependence (Eq. 3). The slight mismatch may be due to the inevitable deviations in the antenna geometry, misalignment and interferences due to the presence of other antennas in the line-of-sight (see suppl. info). We also record the evolution of $V_{opt}$ as a function of $d$, which is shown as the red plot in Fig. 4a. This is done by focusing the laser individually on each antenna and varying $V_{ac}$ to equalize $I_{phot}$ and $I''$ for a sweep of $V_{dc}$ across -0.1 V to 0.1 V(see the supplementary info.). The data are also fitted with a power law with a similar expression. Here $c$ again indicates the noise level, which comes around 6.5 mV. The best fit gives $b= -0.85$ (red solid plot), which is close to -1. This is expected, as $I_{phot}$ is proportional $V_{opt}^2$ (Eq. 2). In the supplementary information, we corroborate the experimental distance dependence with FEM based numerical simulations by calculating the junction's electric field as a function of $d$. As indicated in Eq.3, the amount of power transferred not only depends on the distance, but also on parameters that are influenced by the geometry of the antenna[7,27]. We thus explore the influence of the gap size $g$, and the effect of an off-resonant antenna. These results together with an estimate of the transduction yield of the interconnect are also presented in the supplementary.

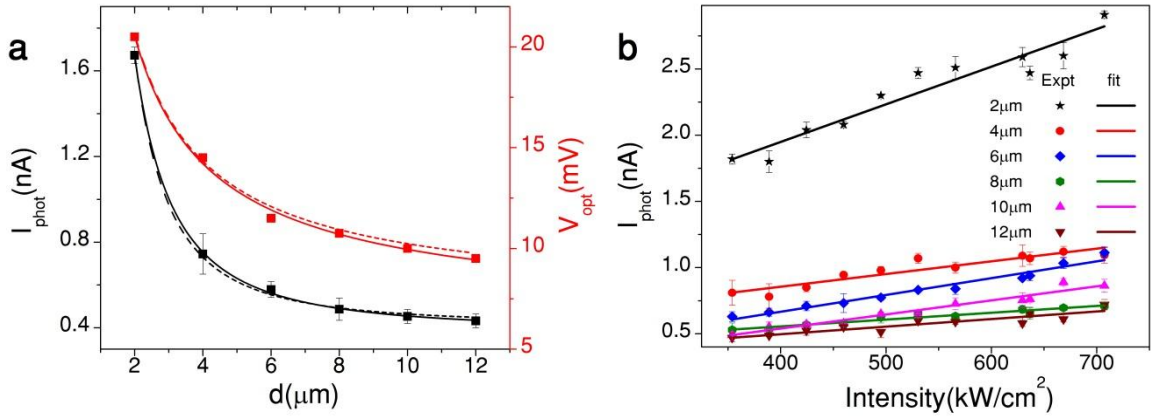

**Figure 4| Characteristics of the optical wireless power transfer. a,** Dependence of $I_{phot}$ and $V_{opt}$ on the distance between the illuminated optical antenna and the rectenna for an excitation intensity of 352 kW.cm$^{-2}$. The solid black and red plot indicates the best fit obtained by using a generic power law function ($\alpha d^b + c$). For comparison, $d^{-2}$ and $d^{-1}$ dependences for $I_{phot}$ and $V_{opt}$ are also shown as black and red dashed plots, respectively. **b,** Plot of $I_{phot}$ as a function of excitation laser intensity for all the antennas. The data are recorded for $V_{dc}$=50 mV.

Finally, we plot in Fig. 4b the dependence of measured rectified photocurrent on the excitation laser intensity for each remote transmitting antenna for $V_{dc}$= 50 mV. Regardless of the distance, the photocurrent scales linearly with the power fed to the optical antennas in agreement with Eq. 3.

In conclusion, we demonstrate an on-chip nanoscale optical wireless link between an laser-illuminated optical antenna and a transducing rectenna. In our experiments, gold nanodisks act as a polarization-sensitive transmitting antenna broadcasting the laser radiation towards the rectifying gap antenna. The amount of power transferred maintains an inverse square relation with the distance between the transmitting antenna and the rectenna. In addition to this, the geometrical properties of the participating units contribute to the yield of power transfer. Further improvement in the efficiency of transmission can be achieved by integration of highly directional optical antennas and a resonant feed. An integrated wireless transmission enables a new communication strategy between nanoscale devices where physical links cannot be implemented. The wireless link can immediately be applied to develop ultrafast optical switches. The plasmonic response of the rectenna results in a large enhancement of the optical field at its atomic scale feed-gap, which makes it an excellent candidate for realizing transistor operations. Transmission rates in excess of $10^{12}$ Mbs$^{-1}$ are achievable as the plasmonic response of the gap is defined by the polarizability response of the metal and not by carrier life time as in semiconductors[28]. Recent progresses show that the electrically driven tunneling junctions can act as ultrafast broadband self-emitting devices[29-31]. Therefore, integration of wireless link between an electrically driven optical antenna with a transducing rectenna may enable ultrafast information broadcasting. Such devices will represent a paradigm for on-chip interfacing of electrons and photons at the nanoscale.

**References**

1    Tse, D. & Viswanath, P. *Fundamentals of wireless communication*. (Cambridge university press, 2005).

2    Chang, M. F. *et al.* CMP Network-on-Chip Overlaid With Multi-Band RF-Interconnect. *IEEE 14th International Symposium on High Performance Computer Architecture* 191-202 (2008)

3    Ruiz-Garcia, L., Lunadei, L., Barreiro, P. & Robla, I. A review of wireless sensor technologies and applications in agriculture and food industry: state of the art and current trends. S*ensors* **9**, 4728-4750 (2009).

4    Varshney, U. Pervasive healthcare and wireless health monitoring. *Mobile Networks and Applications* **12**, 113-127 (2007).

5    Ozbay, E. Plasmonics: merging photonics and electronics at nanoscale dimensions. S*cience* **311**, 189-193 (2006).

6    Alduino, A. & Paniccia, M. Interconnects: Wiring electronics with light. *Nature Photonics* **1**, 153 (2007).

7    Novotny, L. & Van Hulst, N. Antennas for light. *Nature photonics* **5**, 83-90 (2011).

8    Alù, A. & Engheta, N. Wireless at the nanoscale: optical interconnects using matched nanoantennas. *Physical ReviewLetters* **104**, 213902 (2010).

9       Dregely, D. *et al.* Imaging and steering an optical wireless nanoantenna link. *Nature Communications* **5** (2014).

10      Merlo, J. M. *et al.* Wireless communication system via nanoscale plasmonic antennas. *Scientific Reports* **6**, 31710 (2016).

11      Yang, Y., Li, Q. & Qiu, M. Broadband nanophotonic wireless links and networks using on-chip integrated plasmonic antennas. *Scientific Reports* **6** (2016).

12      Solís, D. M., Taboada, J. M., Obelleiro, F. & Landesa, L. Optimization of an optical wireless nanolink using directive nanoantennas. *Optics Express* **21**, 2369-2377 (2013).

13      Ward, D. R., Hüser, F., Pauly, F., Cuevas, J. C. & Natelson, D. Optical rectification and field enhancement in a plasmonic nanogap. *Nature Nanotechnology* **5**, 732-736 (2010).

14      Stolz, A. *et al.* Nonlinear photon-assisted tunneling transport in optical gap antennas. *Nano Letters* **14**, 2330-2338 (2014).

15      Miskovsky, N. M. *et al.* Nanoscale devices for rectification of high frequency radiation from the infrared through the visible: a new approach. *Journal of Nanotechnology* **2012** (2012).

16      Mayer, A., Chung, M., Weiss, B., Miskovsky, N. & Cutler, P. Simulations of infrared and optical rectification by geometrically asymmetric metal–vacuum–metal junctions for applications in energy conversion devices. *Nanotechnology* **21**, 145204 (2010).

17      Du, W., Wang, T., Chu, H.-S. & Nijhuis, C. A. Highly efficient on-chip direct electronic–plasmonic transducers. *Nature Photonics* **11**, 623-627 (2017).

18      Park, H., Lim, A. K., Alivisatos, A. P., Park, J. & McEuen, P. L. Fabrication of metallic electrodes with nanometer separation by electromigration. *Applied Physics Letters* **75**, 301-303 (1999).

19      Mennemanteuil, M.-M. *et al.* Remote plasmon-induced heat transfer probed by the electronic transport of a gold nanowire. *Physical Review B* **94**, 035413 (2016).

20      Tu, X. W., Lee, J. H. & Ho, W. Atomic-scale rectification at microwave frequency. *The Journal of Chemical Physics* **124**, 021105 (2006).

21      Tucker, J. Quantum limited detection in tunnel junction mixers. *IEEE Journal of Quantum Electronics* **15**, 1234-1258 (1979).

22      Tien, P. & Gordon, J. Multiphoton process observed in the interaction of microwave fields with the tunneling between superconductor films. *Physical Review* **129**, 647 (1963).

23      Arielly, R., Ofarim, A., Noy, G. & Selzer, Y. Accurate determination of plasmonic fields in molecular junctions by current rectification at optical frequencies. *Nano Letters* **11**, 2968-2972 (2011).

24      Viljas, J. & Cuevas, J. Role of electronic structure in photoassisted transport through atomic-sized contacts. *Physical Review B* **75**, 075406 (2007).

25      Simmons, J. G. Generalized formula for the electric tunnel effect between similar electrodes separated by a thin insulating film. *Journal of AppliedPphysics* **34**, 1793-1803 (1963).

26      Collin, R. E. *Antennas and radiowave propagation*.  (McGraw-Hill, 1985).

27      Bigourdan, F., Hugonin, J.-P., Marquier, F., Sauvan, C. & Greffet, J.-J. Nanoantenna for electrical generation of surface plasmon polaritons. *Physical Review Letters* **116**, 106803 (2016).

28      Sun, C. *et al.* Single-chip microprocessor that communicates directly using light. *Nature* **528**, 534-538 (2015).

29      Kern, J. *et al.* Electrically driven optical antennas. *Nature Photonics* **9**, 582-586 (2015).

30      Buret, M. *et al.* Spontaneous hot-electron light emission from electron-fed optical antennas. *Nano Letters* **15**, 5811-5818 (2015).

31      ParzefallM *et al.* Antenna-coupled photon emission from hexagonal boron nitride tunnel junctions. *Nature Nanotechnology* **10**, 1058-1063 (2015).

**Acknowledgements**

This work has been supported by the European Research Council under the European Community's Seventh Framework Program FP7/ 2007-2013 Grant Agreement No. 306772. Device fabrication was performed in the technological platform ARCEN Carnot with the support of the Région de Bourgogne. We thank I. Smetanin, O. Demichel for discussions, J. Dellinger for its initial implication in the setup, S. Pernot and B. Sinardet for developping part of the electronic unit.

**Author Contributions**

A.D. conducted the experiment, analyzed the data, and performed the simulations. M.M.M. developed the experimental and fitting procedures, N. C. and M. B. constructed the measurement apparatus to control the electromigration process, G.C.D.F. supervised the simulations. A.B. conceived the experiment and supervised the research. A. D. and A. B. wrote the manuscript, with revisions by all.

**Additional Informations**

Correspondence and requests for materials should be addressed to A.B.

**Competing financial interests**

The authors declare no competing financial interests.

**Methods**

**Sample preparation**

The samples are prepared on a glass coverslip by double step lithography involving electron beam lithography (EBL) and photolithography. First the Au nanodiscs of diameter (220±10 nm), nanoscale constrictions of length 400nm and width 100nm between large triangular Au structures and alignment marks are fabricated by EBL. For this, we spin coat a 250 nm thick double layer PMMA (50kDa and 200kDa) and then sputter a sacrificial gold conductive layer on it to avoid charging during electron beam exposure. Once the PMMA is exposed and developed, a 2nm thick Ti as adhesion layer and 45nm thick Au are subsequently deposited through thermal evaporation. The excess metal is then lifted off to obtain the final nanostructures. Then the macroscopic gold electrodes are designed via standard optical lithography. During this step, alignment marks are used to align the samples coordinate with the coordinate system of the photolithography mask. Once the nanostructrures and electrical connections are fabricated, we produce the nanoscale gap-antenna by controlling the electromigration of the constriction. The electromigration is stopped once the zero bias d.c. conductance reaches to a value lower than quantum conductance ($G_0$=77μS). The presence of a tunnelling gap is confirmed by its nonlinear I-V characteristics.

**Electrical and Optical measurements**

The schematic of the experimental setup we use for the optical and electrical characterizations, presented in the paper is illustrated in the supporting information. We focus the 785nm wavelength laser on the sample through an oil immersion objective lens of numerical aperture (NA) 1.49. During the whole experiment, nanostructures are immersed in refractive index (RI) matching oil of RI=1.52. For electrical measurements, the sample is biased with an applied d.c. bias $V_{dc}$ added with a small modulation a.c. voltage $V_{ac}cos\omega_1 t$ at $V_{ac}$=20mV where $f_1=\omega_1/2\pi$ = 12.37 kHz is the modulation frequency. A first lock in amplifier referenced at $f_1$ and $2f_1$ is used to simultaneously record the first harmonic ($\partial I/\partial V$) and second harmonic ($1/4\ V_{ac}^2\ \partial^2 I/\partial V^2$), of the current tunneling through the gap.

The first harmonic is proportional to the differential conductance and second harmonic signifies nonlinearity in the junction's conductance. To extract the laser-induced current $I_{phot}$ the laser beam is chopped by a chopper at frequency $f_{chop}$= 831 Hz and the tunnelling current is demodulated at $f_{chop}$ using a second lock-in amplifier. Therefore with this arrangement, $I(V_{dc})$, $\partial I/\partial V$, $\partial^2 I/\partial V^2$ and $I_{phot}$ can be measured simultaneously as a function of $V_{dc}$. For mapping, the sample is scanned through the laser spot by the moving the sample with a linearized piezoelectric stage.

# Supplementary Information

# Optical wireless link between a nanoscale antenna and a transducing rectenna

Arindam Dasgupta, Marie-Maxime Mennemanteuil, Mickaël Buret, Nicolas Cazier, Gérard Colas-des-Francs and Alexandre Bouhelier[*]

*Laboratoire Interdisciplinaire Carnot de Bourgogne, CNRS UMR 6303, Université de Bourgogne Franche-Comté, 9 Avenue A. Savary, 21000 Dijon, France.*

[*]e-mail : alexandre.bouhelier@u-bourgogne.fr

## S1. Description of the experimental setup

The experimental setup is built on an inverted microscope (Nikon). A 785nm wavelength laser is focused on the sample through an oil immersion objective lens with a 1.49 numerical aperture (NA). During the whole experiment, the nanostructures are immersed in refractive index (RI) matching oil of RI=1.52. For electrical measurements, the sample is biased with an applied dc bias $V_{dc}$ added with a small modulation ac voltage $V_{ac}cos\omega_{ac}t$ at $V_{ac}$=20mV where $f_{ac}=\omega_{ac}/2\pi$ = 12.37 kHz is the modulation frequency. A first lock in amplifier referenced at $f_{ac}$ and $2f_{ac}$ is used to simultaneously record the 1st harmonic $(\partial I/\partial V)$ and 2nd harmonic $(1/4\ V_{ac}^2\ \partial^2 I/\partial V^2)$, of the current through the gap. The 1st harmonic is proportional to the differential conductance and 2nd harmonic signifies nonlinearity in the junction's conductance. To extract the laser-induced current $I_{phot}$ the laser beam is chopped by a chopper at frequency $f_{chop}$= 831 Hz and the tunnelling current is demodulated at $f_{chop}$ using a second lock-in amplifier. Therefore with this arrangement, $I(Vdc)$, $\partial I/\partial V$, $\partial^2 I/\partial V^2$ and $I_{phot}$ can be measured simultaneously as a function of $V_{dc}$. Also in relevant cases, for monitoring the laser induced change in the differential conductance, we demodulate the $\partial I/\partial V$ signal at the frequency of the chopper $f_{chop}$. For mapping, the sample is scanned through the laser spot by the moving the sample with a linearized piezoelectric stage. Figure S1 schematically represents the experimental arrangement.

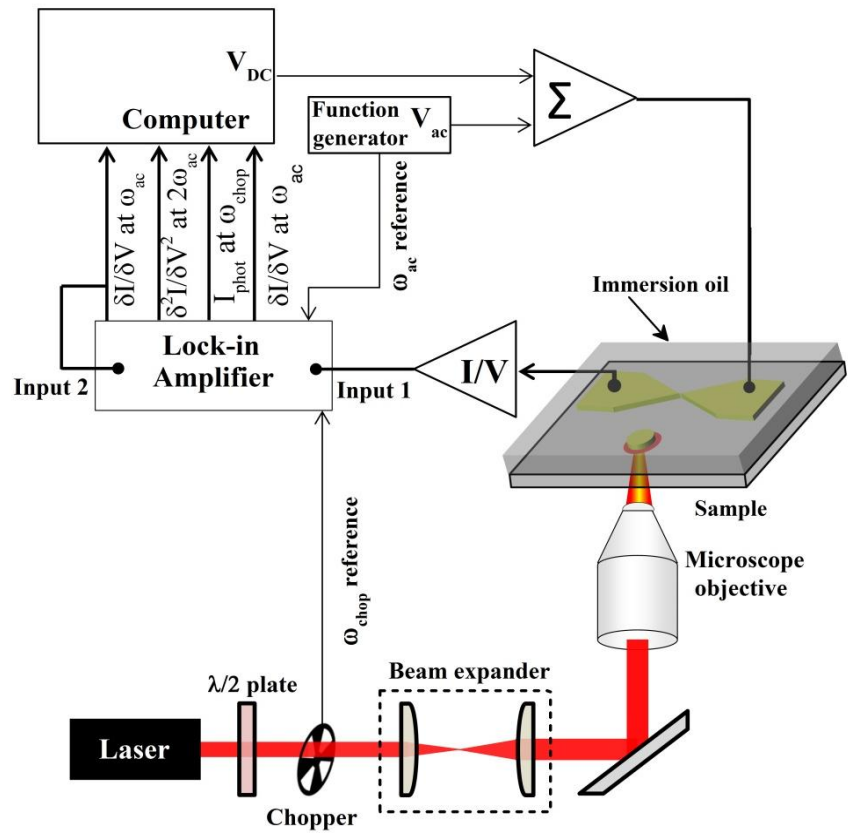

**Figure S1|** Schematic representation of the experimental arrangement for optical and electrical characterization

## S2. Gap size estimation using Simmons' model

We estimate the gap width of the electromigrated tunnel junction by analyzing the experimentally recorded nonlinear I-V characteristics using the general expression to calculate the tunneling current through metal-insultor-metal (MIM) system provided by John G. Simmons.[1,2] Figure S2a is an energy diagram illustrating the electron tunneling through a MIM junction polarized by a bias $V_{dc}$. Here, $\phi_1$, $\phi_2$ are the Schottky barrier heights of the metal on the left side and the right side of the MIM interface and $E_F$ represents the Fermi energy. Following the interpretation provided by Brinkman *et al*, the simplified expression for the tunneling current (*I*) through such a junction is given by,[3]

$$I = A \times \left[ 3.14 \times 10^{-4} \sqrt{\phi} \frac{V_{dc}}{g} - 5.9 \times 10^{-5} \frac{\Delta\phi}{\phi} V_{dc}^2 + 1.64 \times 10^{-4} \frac{g}{\phi} V_{dc}^3 \right] \times \exp\left(-1.025 g \sqrt{\phi}\right) \quad S1$$

Where, $\phi = \frac{\phi_1 + \phi_2}{2}$ and $\Delta\phi = \phi_1 - \phi_2$ in eV represent the average barrier height and the asymmetry in the barrier height on the both side of the MIM junction. In the equation S1, *A* in nm² is the cross-section of the junction and *g* in Å is the gap width.

In our experiment, we expect a very minimal asymmetry in the barrier height over the gap since the material on both sides is similar. However, a residual asymmetry is generally observed[4,5], which probably results from a geometry-dependent modification of the barrier height[6]. In general, for bulk Au-SiO$_2$ interface the height of the Schottky barrier is around 4.5eV as SiO$_2$ has an electron affinity of 0.75eV[7]. However, in case of atomic scale MIM gaps, formation of image charges at the metal-insulator interfaces result in significant lowering of the height of the barrier[8-10]. Likewise, the Ti adhesion layer evaporated between the glass substrate and the Au may form an oxide, which would also reduce the effective barrier height. Based upon a representation of the I-V characteristics in the form of a Fowler-Nordheim plot, Frimmer *et al.* suggested a tunneling transport through the TiO$_2$[11]. However, interpretation of transition voltage spectroscopy as a measure of the barrier height is debated[12].

We include all these aspects into our calculation to estimate all the parameters (g, ϕ and Δϕ) by fitting the experimentally recorded I-V characteristics curve with equation S1 upon fixing the cross section area *A* to a constant value. It is experimentally difficult to infer *A* as electron microscopy provides a general configuration of the junction but failed to indicate where tunneling is really occurring. Figure S2b shows the fitting (red plot) of the experimentally recorded I-V plot (black data points) assuming a cross-section of 100 nm$^2$ (an active area which is 10 nm thick and 10 nm wide) which results in an estimation of the gap width *g*=3.99Å with ϕ=3.05 and Δϕ=-0.59eV. In Fig. S2c, we plot the evolution of all the parameters as a function of various cross section areas. From these calculations, we conclude that the tunneling gap has a width lower than 5 Å.

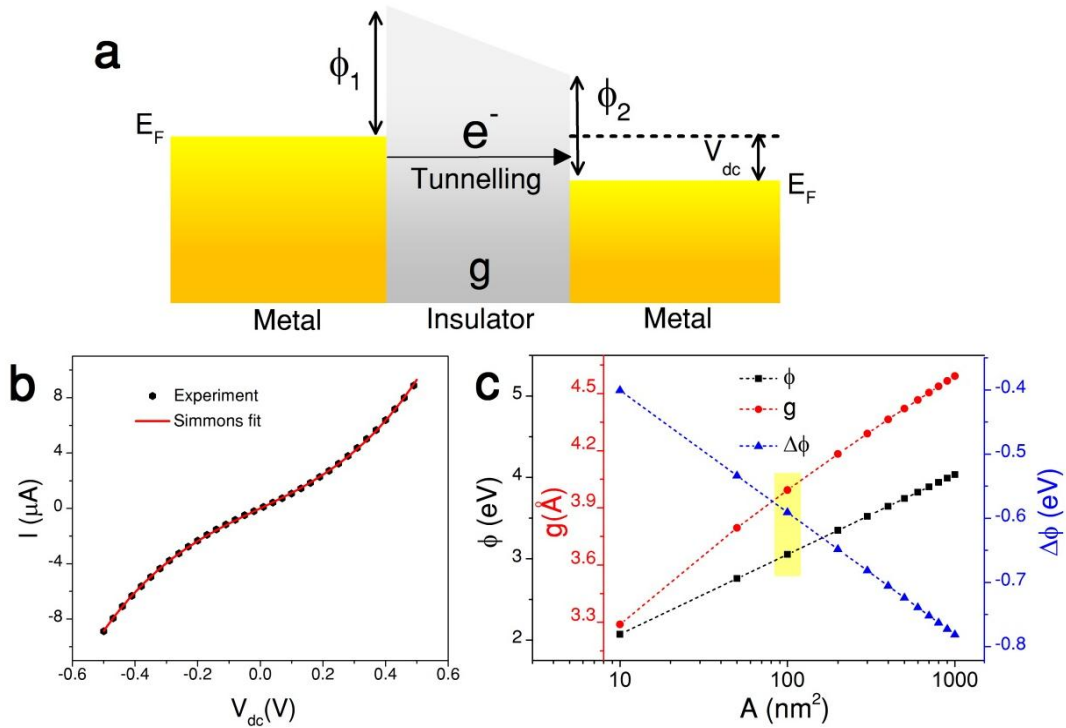

**Figure S2| Estimation of the electromigrated junction gap width. a** Energy diagram illustrating the electron tunneling through an atomic scale MIM junction under applied bias $V_{dc}$. **b** nonlinear I-V characteristics of the electromigrated junction used in our experiment. The black data points represent the experimentally measured I-V characteristics of the junction. The red plot is the Simmons' formula fit of the experimental data with fitting parameters *A*=100 nm$^2$, *ϕ*=3.05 eV, *Δϕ*=-0.59eV and *g*=3.99Å. **c** Evolution of all the fitting parameters (*ϕ*, *Δϕ* and *g*) as a function of cross section area *A*. The points highlighted by yellow color indicate the set of parameters obtained from the fit represented in Fig. S2b.

**S3. Characterization of thermal contributions in optically rectified current**

As discussed in the main manuscript, optical rectification process takes place when the feed-gap receives a radiation, either by a direct illumination or via the mediation of a distant antenna. When the junction is directly illuminated, the photocurrent can be largely dominated by laser-induced thermal effects *i.e.*, thermal expansion of the metallic electrical leads forming the gap and building up of thermo-voltage[13,14] as well as tunneling from photo-excited carriers crossing the barrier height [15]. Laser-induced thermal contributions are expected when the metallic electrodes absorb part of the incoming energy flux and may be observed in $I_{phot}$ map[4,13] even if the feed-gap is outside the excitation area. Let us consider first a laser-induced expansion of the Au electrodes resulting in a reduction of the gap width. The rise in the electrical conductance of the device leads to an increased current flowing through the circuit whenever the laser is positioned on the metallic electrodes. The exact contribution depends on the absorption cross-section of the electrode receiving the incoming light. Furthermore, the absorption of the laser by either of the electrodes creates a temperature gradient across the feed gap leading to a built-up thermo-voltage. However, when the gap is symmetrically illuminated with a centered laser beam, the temperature on both sides is approximately the same and the thermoelectric response is mitigated. Therefore, the effect of a thermo-voltage should be predominant in the $I_{photo}$ map when the laser is positioned on the metal electrode away from the junction.

*Direct illumination of the feed-gap*

To confirm that the measured photocurrent is created through an optical rectification process, we perform a series of experiments to characterize and rule out these thermal effects. First, we record a photocurrent map by scanning the feed-gap through the focused spot at an applied bias $V_{dc}$=0V. As can be seen from the map in Fig. S3a, we observe approximately zero photocurrent when the laser is exactly focused on the gap. This is expected as at $V_{dc}$=0 V, the nonlinearity of the conductance is small and the rectification is thus minimal ($I_{phot}=1/4V_{opt}^2(\partial^2 I/\partial V^2)$) as discussed in main manuscript). Also absence of a response in $I_{photo}$ at the gap suggests that any thermal expansion of the electrodes can be ruled out. Because the metal electrodes are physically strongly bound to the glass surface, thermal expansion is consequently negligible. Furthermore, the tapered geometry of the electrodes acts as a heat sink and is thus more efficient at dissipating the absorbed energy. In the light of the small but measurable asymmetry of the I-V characteristics, the absence of photocurrent when the feed-gap is illuminated also suggests that tunneling of photo-excited carriers above the energy barrier is an unlikely process contributing to the photocurrent. Figure S2a shows an inversion of the sign of $I_{photo}$ when the top or bottom electrode is irradiated by the focused laser. This inversion of the contrast is a clear signature of a thermo-voltage developing across the gap[4,13,16]. These signatures are also present in the $I_{photo}$ maps shown in Fig. S3b and c where applied biases are -50 mV and +50 mV. For nonzero applied biases, we observe an enhancement in $I_{phot}$ response when the laser is positioned on the gap. The fact that the sign of the photocurrent at the gap follows the sign of the applied bias further proves that this is originated through optical rectification. Therefore, we can conclude from this experiment that when the metal-insulator metal junction is illuminated symmetrically with a centered laser beam, only optical rectification is predominant and recorded photocurrent is devoid of any thermal contributions from the laser.

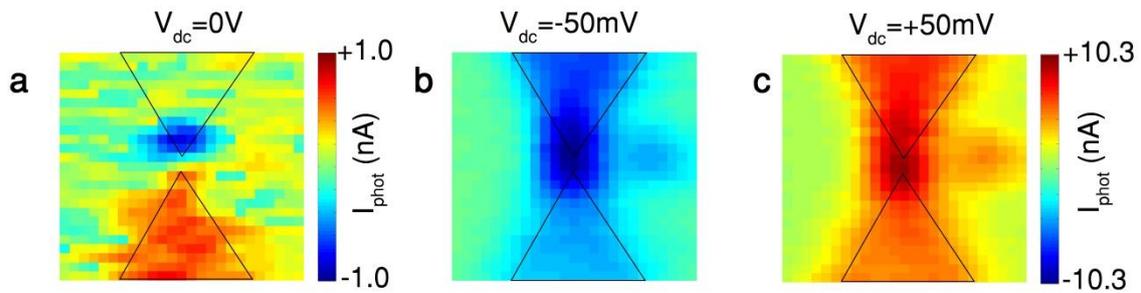

**Figure S3| Photocurrent map for direct exposure of the electromigrated junction. a,** $I_{phot}$ map for a zero applied dc bias ($V_{dc}$=0 V). There is no sign of rectification when the laser is focused on the junction but thermally induced current is visible when laser is focused on one of the electrodes away from the junction. **b,** and **c,** are $I_{photo}$ map of the same area for an applied dc bias of $V_{dc}$=-50 mV and +50 mV respectively. For these, optical rectification is apparent when the laser is focused exactly on the gap. The rectified current follows the sign of the applied d.c. bias.

*Illumination of the transmitter antennas*

To complement the above experiment we monitor the change in $\partial I/\partial V$ at $f_{chop}$ when adjacent nanoantennas are illuminated. Laser-induced thermal variation of the conductance should modulate the recorded photocurrent[13]. We simultaneously map the $I_{phot}$ signal and $\partial I/\partial V$, both at $f_{chop}$ by scanning the laser through the area comprising the optical antennas as presented in Fig. S4a and b, respectively. The incident polarization of the laser is kept along the vertical axis for the entire experiment (maximized rectenna's detection). It is evident from these maps that we could not measure a change even down to $10^{-5}$ level in the conductance map which can be correlated to the recorded $I_{phot}$ signal. Therefore, we can infer from this experiment that the recorded photocurrent when the nanodisks are illuminated is produced through the optical rectification of the transmitted radiation and not due to any laser-induced modulation of the junction conductance.

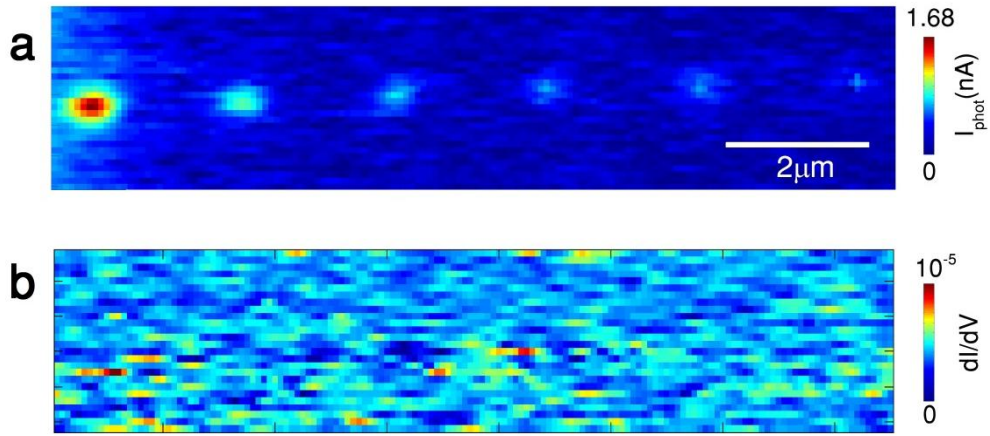

**Figure S4 : a** $I_{phot}$ map presented in Fig. 2a of the main manuscript. **b** Simultaneously acquired map of *dI/dV* demodulated at the frequency of the optical chopper. No measureable contrast in this image can be related to the $I_{phot}$ map. This indicates that the photocurrent are devoided of any laser induced conductance change of the Au feed-gap.

**S4. $V_{opt}$ vs incident polarization measurement.**

To measure the optically induced a.c. voltage drop $V_{ac}$ at the gap, we focus the laser on an individual optical antenna located 4µm away from the junction and simultaneously monitor $I''$ and $I_{phot}$ as a function of applied bias $V_{dc}$ while varying the incident polarization from 0° to 90°. The laser intensity is kept at 540 kW cm$^{-2}$ for the whole experiment. For each incident polarization, $V_{ac}$ is adjusted in such a way so that $I_{phot}$ follows $I''$ for the entire $V_{dc}$ sweep. This is only possible when the optical rectification is the main mechanism behind $I_{phot}$. The results are plotted in Fig. S5. Here data points represent the amplitude of $I_{phot}$ and the lines are $I''$ for the corresponding cases. The value of $V_{ac}$ for which we obtain $I_{phot}=I''$ is recorded as the $V_{opt}$ in the main manuscript.

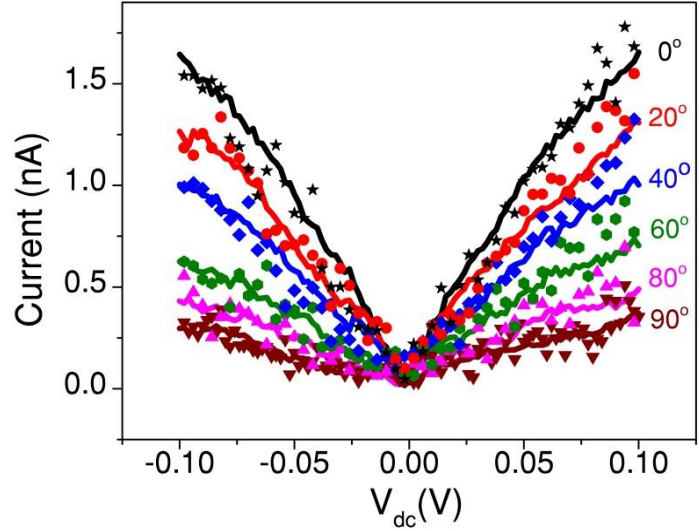

**Figure S5| $V_{opt}$ measurement.** Evolution of $I_{phot}$ and $I''$ for a $V_{dc}$ sweep for different incident polarizations and a laser intensity of 540 kWcm$^{-2}$. The data points represent $I_{phot}$ and the lines represent $I''$ in each plot. For each polarization, $V_{ac}$ is tuned so that $I_{phot}$ and $I''$ are of same amplitude. The value of $V_{ac}$ for conditioning $I_{phot}=I''$ is recorded as $V_{opt}$ in the main manuscript.

**S5. $V_{opt}$ vs distance measurement and transduction yield of wireless link.**

We record the $V_{opt}$ as a function of distance as shown in Fig 4a. of main manuscript by individually focusing the laser on each of the antenna and varying $V_{ac}$ to obtain the situation when $I_{phot}=I''$ for the entire range of the sweep of $V_{dc}$ across -0.1V to +0.1V. The experimental data for three nanodisks are shown in Fig S6. The transduction yield increases as we increase the applied d.c. bias to the junction. This is expected as the nonlinearity of the conductance of the rectenna increases as we raise $V_{dc}$.

We also estimate the transduction yield, which is a measure of what fraction of the incident laser power is converted to electrical power through rectification. The transduction yield $\phi$ in dBm can be given by the following equation,

$$\phi \text{ (in dBm)} = 10 \times \log\left(\frac{I_{phot}^2}{G \times (1mW \text{ of incident power})}\right) \qquad S1$$

where $G$ is the conductance of the MIM junction.

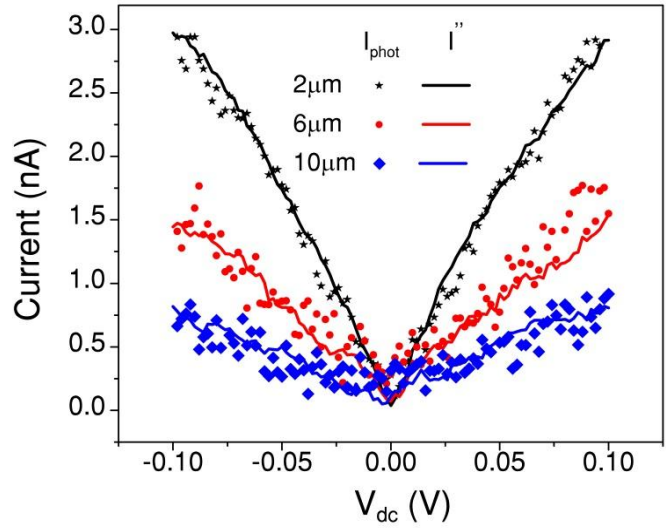

**Figure S6 |** Variation of $I_{phot}$ and $I''$ as a function of $V_{dc}$ for three distant optical antennas

We estimate $\phi$=-91 dBm for a radiation transmitted by the antenna situated 2 μm away and $V_{dc}$= 50mV. Raising $V_{dc}$ to 100mV improves the transduction yield to -87 dBm. Here $\phi$ is the overall efficiency of the link including signal propagation and transduction. In radio-frequency wireless communication, the quality of the transmitted signal is evaluated by the received signal strength indicator (RSSI) measured at the reception node before transduction. RSSI comprised between -70 dBm to -100 dBm benchmark the transmission channel as good to fair.

## S6. Numerical simulations of distance and geometrical parameter dependence of the junction electric field at the rectenna feed-gap.

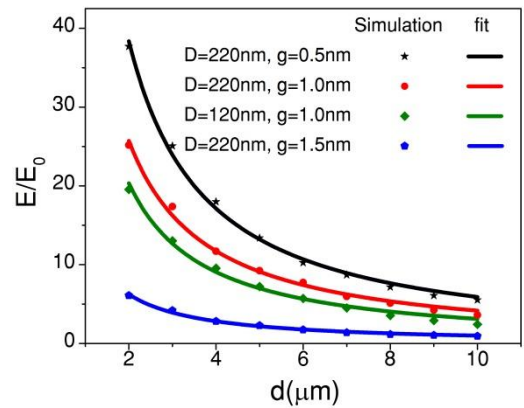

**Figure S7| Numerical simulation of the distance and geometrical parameter dependence of the junction electric field at the rectenna feed gap.** Simulated results of the electric field amplitude at the feed-gap vs d for different combinations of gap widths and antenna diameters. The power law fits (line plots) always converge to an exponent close to -1. The field amplitude decreases as we increase the gap width and an off-resonant antenna (D=120nm) instead of the resonant one (D=220nm).

We corroborate our experimental distance dependence of the rectified current and optically induced voltage drop at the junction (Fig. 4a of main manuscript) with 3 dimension- finite element method (3D-FEM) based calculation of the electric field created at the rectenna's feed-gap for varying distance $d$ between the rectenna and the nanodisk. The normalized electric field produced at a gap size of 0.5 nm and a 220 nm transmitter antenna is presented in Fig. S7 (black data points). The power law fit to the distance dependence (black line) converges to an exponent close -1, which supports the experimental dependence of $V_{opt}$.

As indicated in Eq. 3 in the main manuscript, the amount of power transferred not only depends on the distance, but also on parameters that are influenced by the geometry of the antenna[17,18]. We thus explore the influence of the gap size $g$, and the effect of an off-resonant antenna. Fig. S7 indicates that the electric field at the feed-gap is larger for sub-nm separation. However, our calculations do not take into account quantum-size effect[19], which may limit the amplitude of the electrical field. Broadcasting the optical signal with a non-resonant antenna of diameter D=120 nm also leads to a reduced electrical field at the transducing feed-gap. The presence of other nanodisks in the path transmission is minimal and discussed in the following section.

## S7. Effect of the presence of other antennas in the line-of-sight.

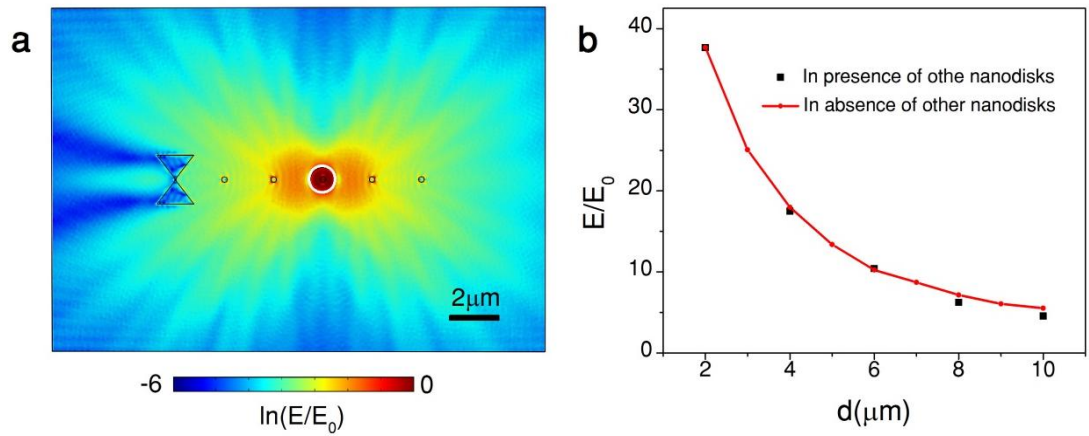

**Figure S8| Effect of the presence of adjacent antennas in the path of transmission. a** The electric field distribution around the nanostructures when a nanodisk at a distance of 6μm is excited in presence of other nanodisks in the transmission path. **b** The red plot represents the electric field values derived when the calculation does not include any antennas placed in line-of-sight. The black data points represents the calculated electric field values when 5 nanodisks are always present in the simulation geometry and each of them is excited individually.

In this section, we numerically estimate the effect of the presence of other nanodisks in the path of transmission towards the rectenna on the field localization at the rectenna feed-gap. For that, we place five antennas at incremental distances from the rectenna with a step size of 2 μm. We calculate the normalized electric field at the junction by illuminating one of these antennas at a time with the vertical incident polarization. In Fig. S8a, we plot the calculated field distribution around the structures in logarithmic scale when an optical antenna located 6 μm away from the feed-gap is excited. It is clear that the presence of the nanoparticles induces additional scattering of the field radiated from the excited element. In the plot of Fig. S8b, we compare the calculated electric field values (black data points) with the values (red plot) determined when the calculations do not include other disks in the line of sight (Fig. S7). The electric field decreases slightly and at maximum 17% reduction is observed for the disc which is 10μm away from the junction. The transmitted radiation is thus shadowed due to the presence of scattering elements in the path to the rectenna, but the effect is minimal in this scenario.

**References**

1  Simmons, J. G. Generalized formula for the electric tunnel effect between similar electrodes separated by a thin insulating film. *Journal of Applied Physics* **34**, 1793-1803 (1963).

2  Simmons, J. G. Electric tunnel effect between dissimilar electrodes separated by a thin insulating film. *Journal of Applied Physics* **34**, 2581-2590 (1963).

3  Brinkman, W., Dynes, R. & Rowell, J. Tunneling conductance of asymmetrical barriers. *Journal of Applied Physics* **41**, 1915-1921 (1970).

4  Stolz, A. *et al.* Nonlinear photon-assisted tunneling transport in optical gap antennas. *Nano Letters* **14**, 2330-2338 (2014).

5  Buret, M. *et al.* Spontaneous hot-electron light emission from electron-fed optical antennas. *Nano Letters* **15**, 5811-5818 (2015).

6  Mayer, A. *et al.* Analysis of the efficiency with which geometrically asymmetric metal–vacuum–metal junctions can be used for the rectification of infrared and optical radiations. *Journal of Vacuum Science & Technology B, Nanotechnology and Microelectronics: Materials, Processing, Measurement, and Phenomena* **30**, 031802 (2012).

7  Nobuyuki, F., Akio, O., Katsunori, M. & Seiichi, M. Evaluation of valence band top and electron affinity of SiO 2 and Si-based semiconductors using X-ray photoelectron spectroscopy. *Japanese Journal of Applied Physics* **55**, 08PC06 (2016).

8  Binnig, G., Garcia, N., Rohrer, H., Soler, J. & Flores, F. Electron-metal-surface interaction potential with vacuum tunneling: Observation of the image force. *Physical Review B* **30**, 4816 (1984).

9  Ma, X., Shu, Q., Meng, S. & Ma, W. Image force effects on trapezoidal barrier parameters in metal–insulator–metal tunnel junctions. *Thin Solid Films* **436**, 292-297 (2003).

10  Nguyen, H., Q., Feuchtwang, T., E. & Cutler, P., H. Do tunneling electrons probe the image interaction? *Journal de Physique Colloques* **47**, C2-37-C32-44 (1986).

11  Frimmer, M., Puebla-Hellmann, G., Wallraff, A. & Novotny, L. The role of titanium in electromigrated tunnel junctions. *Applied Physics Letters* **105**, 221118 (2014).

12	Vilan, A., Cahen, D. & Kraisler, E. Rethinking transition voltage spectroscopy within a generic Taylor expansion view. *ACS Nano* **7**, 695-706 (2012).

13	Ward, D. R., Hüser, F., Pauly, F., Cuevas, J. C. & Natelson, D. Optical rectification and field enhancement in a plasmonic nanogap. *Nature Nanotechnology* **5**, 732-736 (2010).

14	Zolotavin, P., Evans, C. & Natelson, D. Photothermoelectric effects and large photovoltages in plasmonic Au nanowires with nanogaps. *The Journal of Physical Chemistry Letters* **8**, 1739-1744 (2017).

15	Diesing, D., Merschdorf, M., Thon, A. & Pfeiffer, W. Identification of multiphoton induced photocurrents in metal–insulator–metal junctions. *Applied Physics B* **78**, 443-446 (2004).

16	Xu, X., Gabor, N. M., Alden, J. S., van der Zande, A. M. & McEuen, P. L. Photo-thermoelectric effect at a graphene interface junction. *Nano Letters* **10**, 562-566 (2009).

17	Novotny, L. & Van Hulst, N. Antennas for light. *Nature Photonics* **5**, 83-90 (2011).

18	Bigourdan, F., Hugonin, J.-P., Marquier, F., Sauvan, C. & Greffet, J.-J. Nanoantenna for electrical generation of surface plasmon polaritons. *Physical Review Letters* **116**, 106803 (2016).

19	Esteban, R., Borisov, A. G., Nordlander, P. & Aizpurua, J. Bridging quantum and classical plasmonics with a quantum-corrected model. *Nature Communications* **3**, 825 (2012).